# Highly-Efficient Single-Switch-Regulated Resonant Wireless Power Receiver with Hybrid Modulation

Kerui Li, *Student Member, IEEE*, Albert Ting Leung Lee, *Member, IEEE,* Siew-Chong Tan, *Senior Member, IEEE,* and Ron Shu Yuen Hui, *Fellow, IEEE*

***Abstract*—In this paper, a highly-efficient single-switch-regulated resonant wireless power receiver with hybrid modulation is proposed. To achieve both high efficiency and good output voltage regulation, phase shift and pulse width hybrid modulation are simultaneously applied. The soft switching operation in this topology is achieved by the cycle-by-cycle phase shift adjustment between the input current and the gate drive signal and also attributed to the reactive components such as the series-compensated secondary coil ($L_s$, $C_s$) and the parasitic capacitor of the active switch ($C_{s1}$). The soft switching operation also leads to high efficiency and low EMI. By adjusting the duty ratio of the switch, tight regulation of the output voltage can be attained. The steady-state and dynamic models of the resonant receiver with hybrid modulation are analytically derived in order to properly design the feedback controller. An experimental setup of a two-coil wireless power transfer (WPT) system, including the hardware prototype of the proposed receiver, is constructed for experimental verification. The experimental results show the effectiveness of the soft-switching operation in the receiver with a maximum AC−DC efficiency of 98% while maintaining good regulation of the output voltage, regardless of line and load variations.**

## I. Introduction

The rapid advancement in wireless power technology has completely transformed the way we recharge a myriad of battery-powered portable electronic devices because of its added convenience, durability and safety [1]−[6]. Nowadays, an increasing number of consumer electronic devices such as smartphones, smart watches, headsets, and tablets have already incorporated the Qi wireless charging function. According to the latest market survey [7], the total shipment of wireless power receivers is more than doubled that of the wireless power transmitter by 2022. To cope with the fast-growing market demand and more stringent requirements in medium-power WPT system [8], it is advantageous to develop a wireless power receiver that is efficient, compact, low-cost, and reliable for practical applications. Hence, the motivation of this research work is the introduction of a highly-efficient single-switch-regulated resonant power receiver with hybrid modulation, which is characterized by a simple circuit structure, low component count, high AC-DC conversion efficiency, and good output voltage regulation.

The conventional two-stage topology, which comprises a diode bridge rectifier and a regulated DC−DC converter (or LDO linear regulator), is more prevalent in commercial wireless power receivers [9]−[15]. Compared with the single-stage counterpart, the two-stage solution is bulky, costly, inefficient and less reliable because it requires more power switches and passive components. In light of this, considerable research efforts have been devoted to developing various types of single-stage solutions [16]−[21]. In particular, an active full-bridge rectifier is introduced, which can concurrently achieve high-frequency rectification and output voltage regulation [21]. Unfortunately, no soft switching operation is allowed, which degrades its efficiency and EMI performance. In addition, the synchronization between the resonant input current and gate driving signals becomes more sophisticated at higher switching frequencies, which may not be suitable for high-frequency WPT applications. In [19], even though the half-bridge-based solution can attain good output regulation, the use of complementary high-low-side active switches on the same bridge leg increases the risk of a direct short between the DC bus voltage and ground, which unavoidably undermines the reliability and robustness of the receiver. Moreover, the use of pulse density modulation produces significant output voltage ripples. A relatively bulky output capacitor is therefore needed to mitigate the output ripples.

Recently, a number of single-stage single-switch wireless power receivers have emerged [16], [17], [22], which employ fewer power switches, thus making them more attractive solutions for WPT by offering higher reliability and lower cost. Nonetheless, these prior art have their own limitations. For example, the duty-cycle-regulated class-E rectifier in [16] suffers from high voltage stresses on the main switch and the diodes, which inevitably degrades its reliability and potentially shortens its operating lifetime. In [17], the phase-shift regulated class-E rectifier requires complicated design procedures and suffers from poor efficiency at light load condition. In [22], the main disadvantages of the multi-cycle-switching-regulated active rectifier include the use of a bulky output capacitor, a narrow regulation range, and lower efficiency.

In this paper, a single-switch-regulated resonant wireless power receiver is proposed. This wireless power receiver carries the following merits: 1) concurrent high-frequency AC−DC rectification and DC regulation; 2) a wide range of output regulation; 3) robust output regulation against line and load disturbances; 4) soft-switching operation regardless of the coupling and load conditions; and 5) highly-efficient AC−DC energy conversion. This paper is organized as follows. Section II presents the circuit topology of the proposed single-switch resonant wireless power receiver and its operating principle. Section III discusses the design consideration, particularly, how to determine the minimum value of the output capacitor in order to achieve a nearly-constant output voltage. Section IV compares the proposed wireless power receiver with its predecessors. Section V contains the experimental results. Finally, Section VI concludes this research work.

## II. Highly-Efficient Single-Switch Regulated Resonant Wireless Power Receiver System

### A. Circuit Topology

Fig. 1 shows a generic two-coil WPT system, with the emphasis on the proposed wireless power receiver. On the transmitter side, it is assumed that a voltage source inverter (VSI) (e.g. full-bridge/half-bridge inverter), that generates a square waveform with a fundamental frequency of $f_s$ and a series-compensated primary coil ($L_p$, $C_p$), where $f_s = 1/T_s = 1/2\pi\sqrt{C_pL_p}$, is used. On the other hand, the wireless power receiver is made up of the series-compensated secondary coil ($L_s$, $C_s$, and $f_s = 1/2\pi\sqrt{C_sL_s}$) and the single-switch resonant rectifier circuit, which consists of a power switch ($S_1$), three capacitors ($C_{S1}$, $C_{D1}$, $C_o$), a diode ($D_1$), a resistive load ($R$), a synchronization circuit, and a microcontroller (MCU).

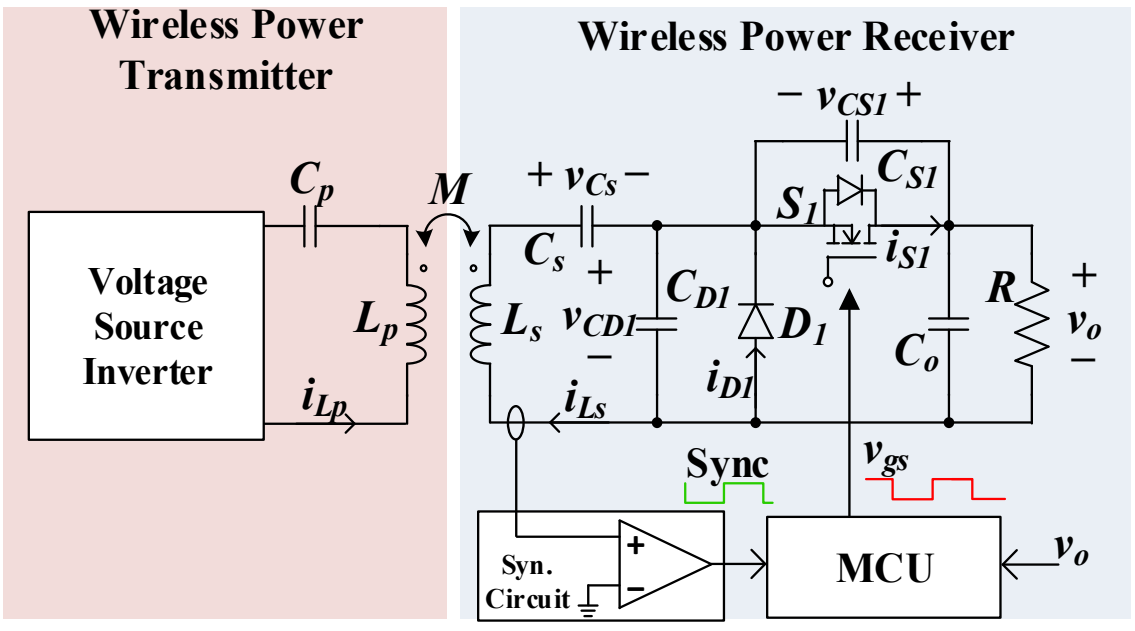

Fig. 1. Schematic diagram of a generic two-coil WPT system with the proposed wireless power receiver.

Basically, the topology of the proposed receiver is a semi-active class-D rectifier. $S_1$ and $D_1$ operate in a complementary manner which determines the nominal value of the output voltage. $C_{S1}$ and $C_{D1}$ are parasitic capacitors of $S_1$ and $D_1$, respectively. $C_o$ is the output capacitor for minimizing the output voltage ripple.

In contrast to conventional class-D synchronous rectifiers for DC-DC conversion applications [23]–[25], the proposed wireless power receiver overcomes the following design challenges: 1) Enable synchronization between resonant current and the gate driving waveform of the receiver; 2) Maintain soft switching and good output regulation for a wide range of load and coupling coefficient values by employing hybrid modulation; 3) Make use of standard steady-state and small-signal models to analyze static and dynamic performance of the closed-loop system; and 4) Achieve concurrent high-frequency AC−DC rectification and DC output regulation by employing the model-based controller design method.

### B. Operating Principles

In the ensuing discussion, the following assumptions are made: 1) For simplicity, ideal circuits are assumed and hence, the equivalent series resistance (ESR) of the passive component can be neglected; 2) The quality factor of the resonant tanks, namely, ($L_p$, $C_p$) and ($L_s$, $C_s$), is sufficiently high so that only the fundamental component of the current is taken into consideration; and 3) The value of the output capacitor $C_o$ is large enough so that a constant DC voltage with reasonably small ripple is produced.

Due to the inherent property of series-series compensation, the input current of the wireless power receiver $i_{Ls}$ is only a function of the fundamental component of the output voltage of the VSI at the transmitter side and the coupling coefficient $M$ [26], [27]. Consequently, $i_{Ls}$ can be treated as a current source at the receiver side, which is expressed as

$$i_{Ls} = |\mathrm{I}_{Ls}|\sin(2\pi f_s t). \quad (1)$$

By using $i_{Ls}$ as a reference signal, the ideal waveforms of the key signals (i.e., $v_{gs}$, $v_{CS1}$, $v_{CD1}$, $i_{CD1}$, and $v_o$) of the single-switch-regulated resonant wireless power receiver wireless are graphically illustrated in Fig. 2.

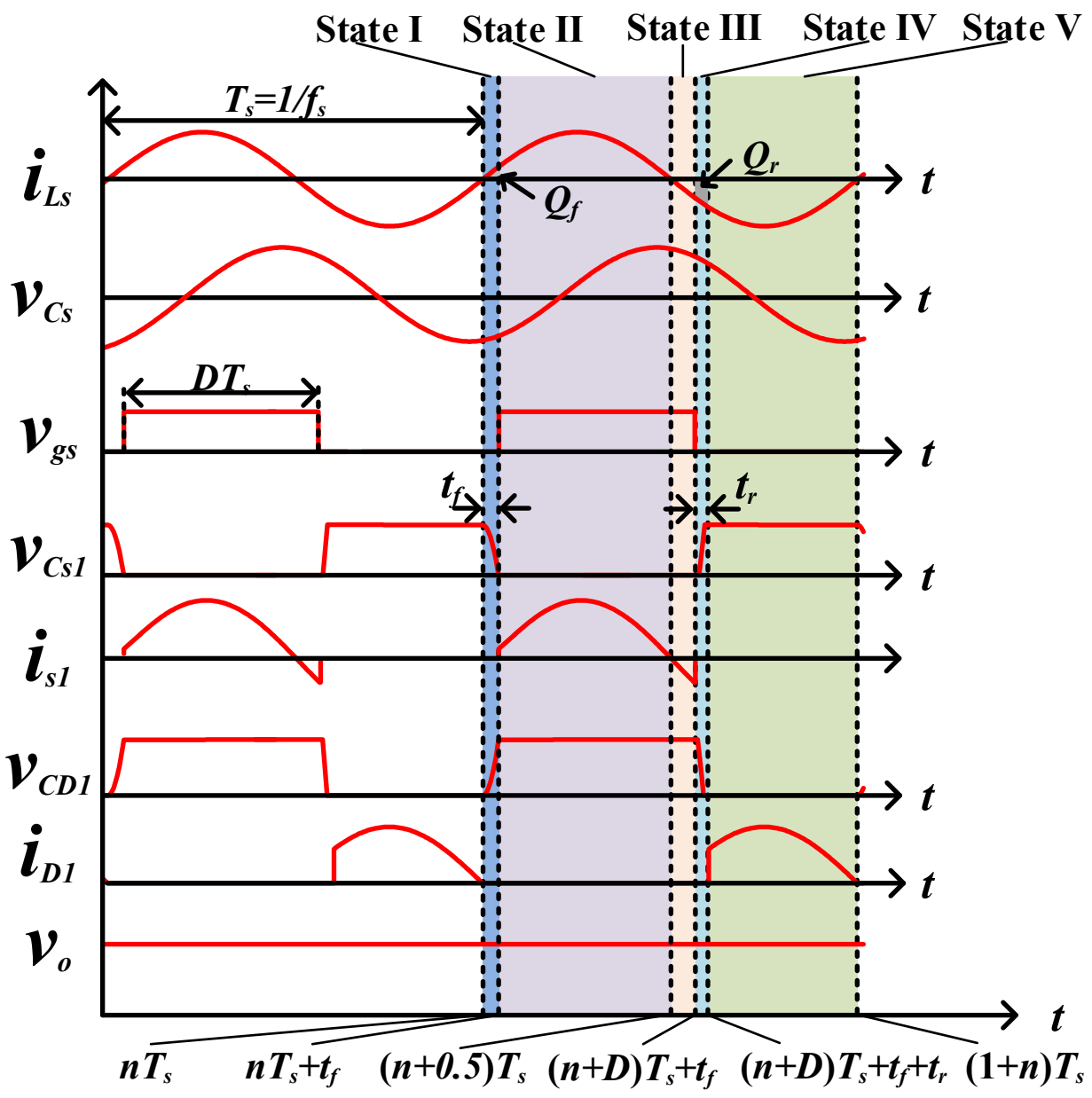

Fig. 2. Key waveforms of the single-switch-regulated resonant wireless power receiver system.

In particular, the gate-to-source voltage ($v_{gs}$) of $S_1$, i.e., the gate drive PWM signal (with a duty ratio of $D$), has a time delay of $t_f$ relative to the zero-crossing point of the positive cycle of $i_{Ls}$. By considering the on/off switching state of $S_1$ and the charging/discharging period of $C_{S1}$ and $C_{D1}$, four distinct states can be defined for the receiver, namely, State I for $nT_s \le t < (nT_s + t_f)$, State II for $(nT_s + t_f) \le t < (n + D)T_s + t_f$, State III for $(n + D)T_s + t_f \le t < (n + D)T_s + t_f + t_r$, and State IV for $(n + D)T_s + t_f + t_r \le t < (1 + n)T_s$, where $t_f$ and $t_r$ are the fall and rise time of the voltage across $C_{S1}$ (i.e., $V_{CS1}$), respectively, $n$ is an arbitrary nonnegative integer number, and $D$ is the on-time duty ratio of the switch. Fig. 3(a)–(d) shows the equivalent circuit model in each of the four states.

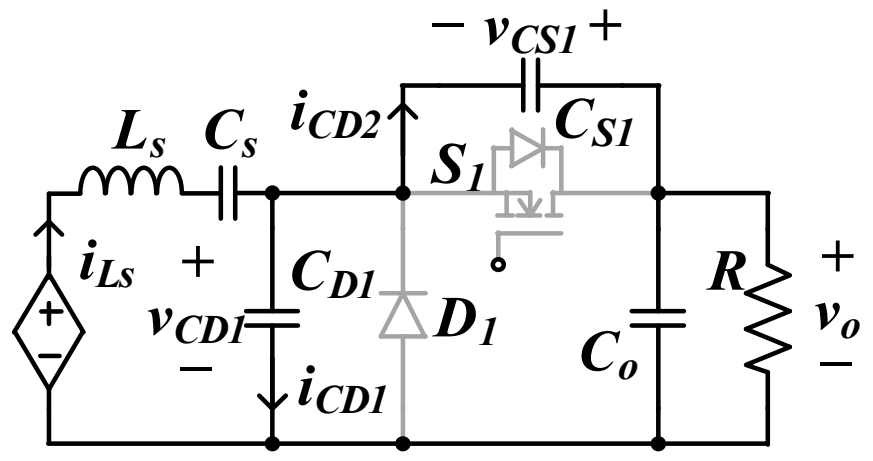

(a) State I

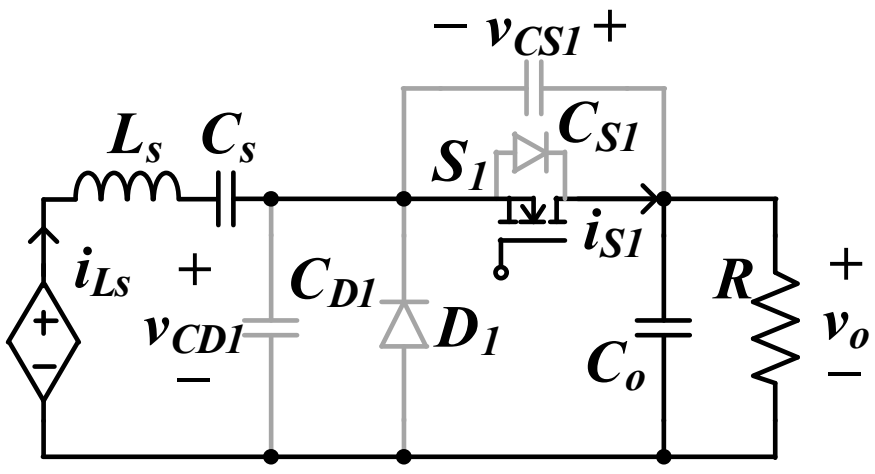

(b) State II

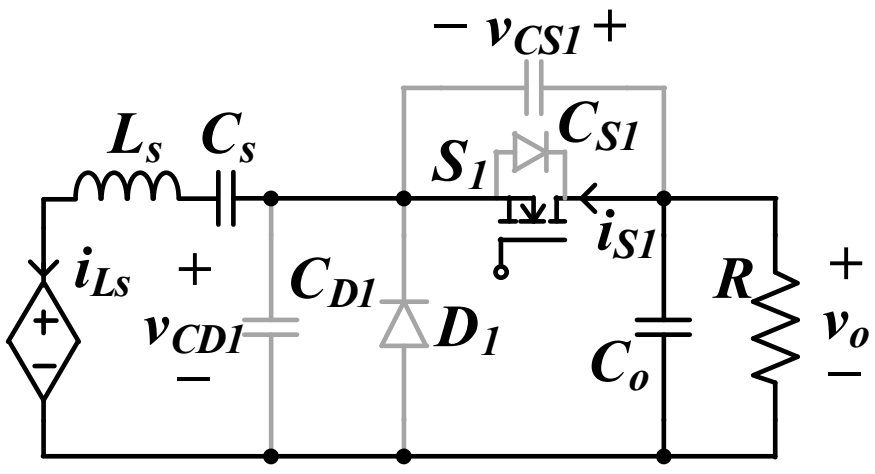

(c) State III

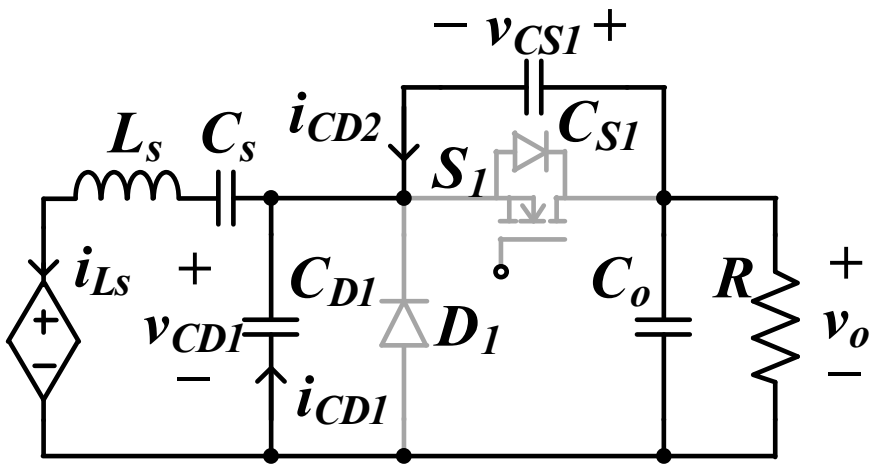

(d) State IV

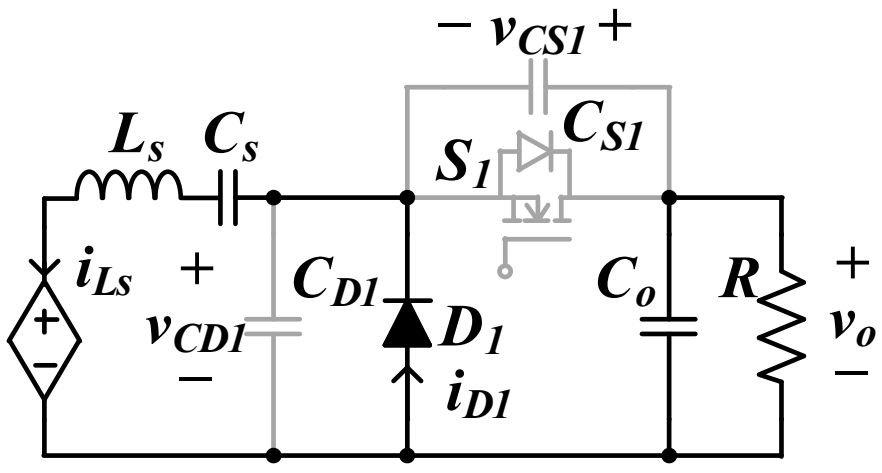

(e) State V

Fig. 3 Equivalent circuit model in (a) State I; (b) State II; (c) State III; (d) State IV, (e) State V.

State I [$nT_s \leq t < (nT_s + t_f)$]:
Fig. 3(a) shows the equivalent circuit model of the receiver circuit in State I. At $t = nT_s$, the diode current $i_{CD1}$ drops to zero and stops conducting naturally, resulting in a zero-current-switching (ZCS) turn-off operation of the diode. When $t > nT_s$, $i_{Ls}$ charges $C_{D1}$ and also discharges $C_{S1}$ at the same time. By using KCL and KVL, we have

$$\begin{cases} C_{D1}\dfrac{dv_{CD1}}{dt} - C_{S1}\dfrac{dv_{CS1}}{dt} = i_{Ls} \\ v_{CD1} + v_{CS1} = v_O. \end{cases} \quad (2)$$

Meanwhile, the output capacitor $C_o$ supplies current to the load $R$. Mathematically, we can write

$$C_O\frac{dv_o}{dt} = -\frac{v_o}{R}. \quad (3)$$

State II [$(nT_s + t_f) \leq t < (n + 0.5)T_s$]:
Fig. 3(b) contains the equivalent circuit model of the receiver circuit in State II. At $t = (nT_s + t_f)$, $v_{Cs1}$ returns to zero volts and afterwards, $S_1$ is turned on at the rising edge of $v_{gs}$. This enables zero-voltage-switching (ZVS) turn-on operation. After $S_1$ is switched on, $v_{CS1}$ and $v_{CD1}$ are clamped at zero volts and $v_o$, respectively, i.e., $v_{Cs1}(nT_s + t_f) = 0$ and $v_{CD1}(nT_s + t_f) = v_o$. By substituting these two conditions into (2) and solving for $t_f$, where $t_f$ is the time interval of State I ($t_f$), we have

$$t_f \approx \sqrt{\frac{(C_{D1} + C_{S1})v_O}{\pi f_s |\mathrm{I}_{Ls}|}}. \quad (4)$$

State III [$(n + 0.5)T_s \leq t < (n + D)T_s + t_f$]:
Fig. 3(c) contains the equivalent circuit model of the receiver circuit in State II. At $t = (n+0.5)T_s$, $i_{Ls}$ returns to zero and becomes negative, while $S_1$ remains on. During State II and III, $i_{Ls}$ transfers the input energy to both the output capacitor $C_o$ and the load $R$ via $S_1$, which can be written as

$$C_O\frac{dv_o}{dt} = i_{Ls} - \frac{v_o}{R}. \quad (5)$$

To allow proper regulation of the output voltage ($v_o$), the range of the duty ratio ($D$) is defined as

$$\frac{1}{2} - f_s t_f \leq D \leq 1 - 2f_s t_f. \quad (6)$$

State IV [$(n + D)T_s + t_f \leq t < (n + D)T_s + t_f + t_r$]:
Fig. 3(d) shows the equivalent circuit model of the receiver circuit in State III. At $t = (n + D)T_s + t_f$, $S_1$ is turned off with zero voltage. Afterwards, $i_{Ls}$, which is in the negative half cycle, starts discharging $C_{D1}$ and charging $C_{S1}$ simultaneously. The charging procedure can thus be mathematically represented as

$$\begin{cases} C_{D1}\dfrac{dv_{CD1}}{dt} - C_{S1}\dfrac{dv_{CS1}}{dt} = i_{Ls} \\ v_{CD1} + v_{CS1} = v_O \end{cases} \quad (7)$$

Meanwhile, $C_o$ continues to supply current to the load $R$. Hence, we have

$$C_O\frac{dv_o}{dt}=-\frac{v_o}{R}. \qquad (8)$$

State V $[(n + D)T_s + t_f + t_r \le t < (1 + n)T_s]$:

Fig. 3(e) contains the equivalent circuit model of the receiver circuit in State IV. At $t = (n + D)T_s + t_f + t_r$, $v_{CD1}$ returns to zero volts and $D_1$ starts conducting. After $D_1$ is forward biased, $v_{CD1}$ and $v_{CD1}$ are clamped at $v_o$ and zero volts, respectively, i.e., $v_{CS1}[(n + D)T_s + t_f + t_r] = v_o$ and $v_{CD1}[(n + D)T_s + t_f + t_r] = 0$. By substituting these two conditions into (6), $t_r$ can be obtained as

$$t_r=\frac{1-D}{f_s}-t_f-\frac{1}{2\pi f_s}\arccos\left(\cos\left(2\pi(D+f_st_f)\right)+\frac{2\pi f_s(C_{D1}+C_{S1})v_o}{|\mathrm{I}_{Ls}|}\right). \qquad (9)$$

From Fig. 2, $Q_f$ and $Q_r$ represent the total charge transferred during the switching intervals $t_f$ and $t_r$, respectively. Since $Q_f = Q_r$, the average current of $i_{LS}$ in State III is higher than that in State I. Hence, $t_f$ becomes larger than $t_r$. Note that in this particular state, when $D_1$ conducts, $i_{Ls}$ freewheels through $D_1$ and no energy is transferred to the output. $C_o$ continues to discharge its current to the load $R$. By KCL, we can write

$$C_O\frac{dv_o}{dt}=-\frac{v_o}{R}. \qquad (10)$$

At $t = (1 + n)T_s$, the circuit enters State I of the next switching period and the aforementioned state transitions will repeat.

*C. Steady-State Model*

By invoking state-space averaging on the output capacitor $C_o$, the relationship between the duty ratio ($D$) and the output voltage ($v_o$) based on the steady-state model can be analytically derived as

$$C_O\frac{dv_o}{dt}=\frac{1}{T_s}\int_{nT_s+t_f}^{(n+D)T_s+t_f} i_{Ls}\,dt-\frac{v_o}{R}$$
$$=\frac{|\mathrm{I}_{Ls}|}{2\pi}\left(\cos(2\pi f_st_f)-\cos(2\pi D+2\pi f_st_f)\right)-\frac{v_o}{R}. \qquad (11)$$

Since $\frac{dv_o}{dt}=0$ in steady-state condition, the L.H.S. of (11) becomes zero and hence, by re-arranging, $v_o$ can be obtained as

$$v_o=\frac{|I_{Ls}|R}{2\pi}\left(\cos(2\pi f_st_f)-\cos(2\pi D+2\pi f_st_f)\right). \qquad (12)$$

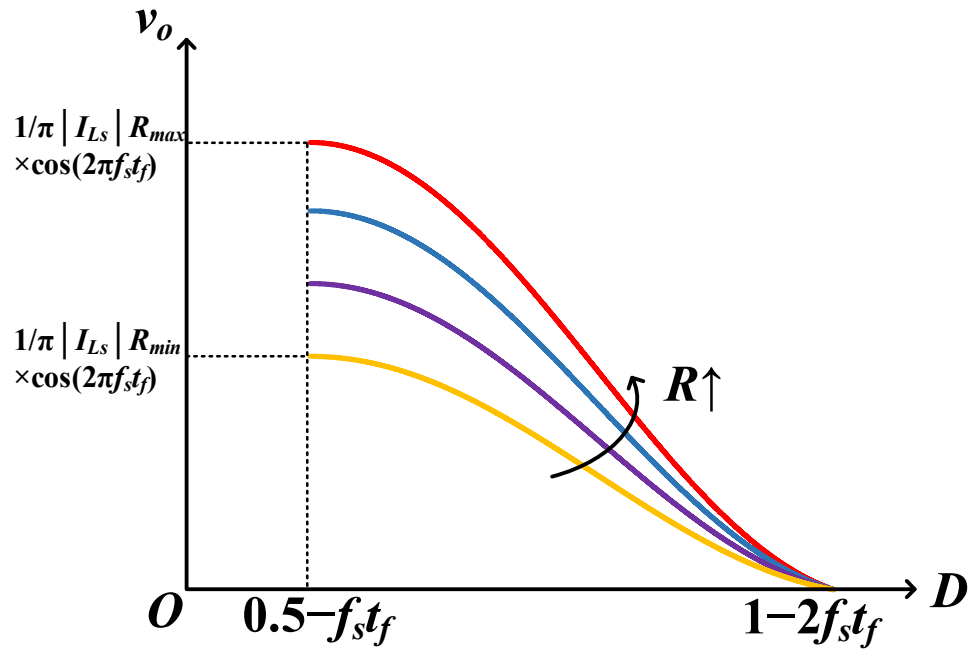

(a) Plots of output voltage $v_o$ versus duty cycle $D$ at different values of $R$.

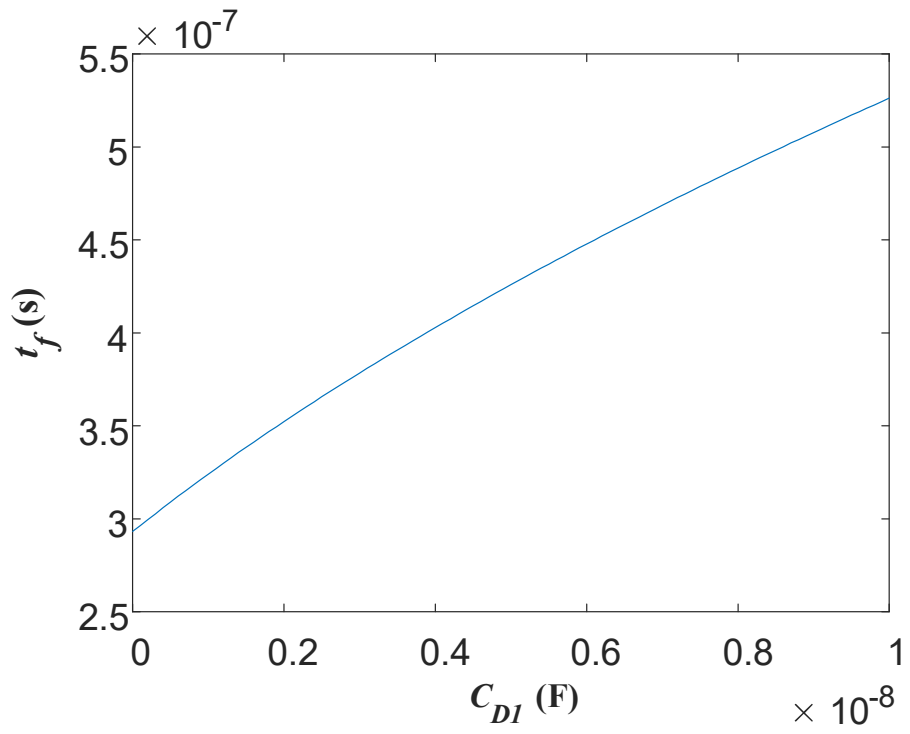

(b) $t_f$ versus $C_{D1}$

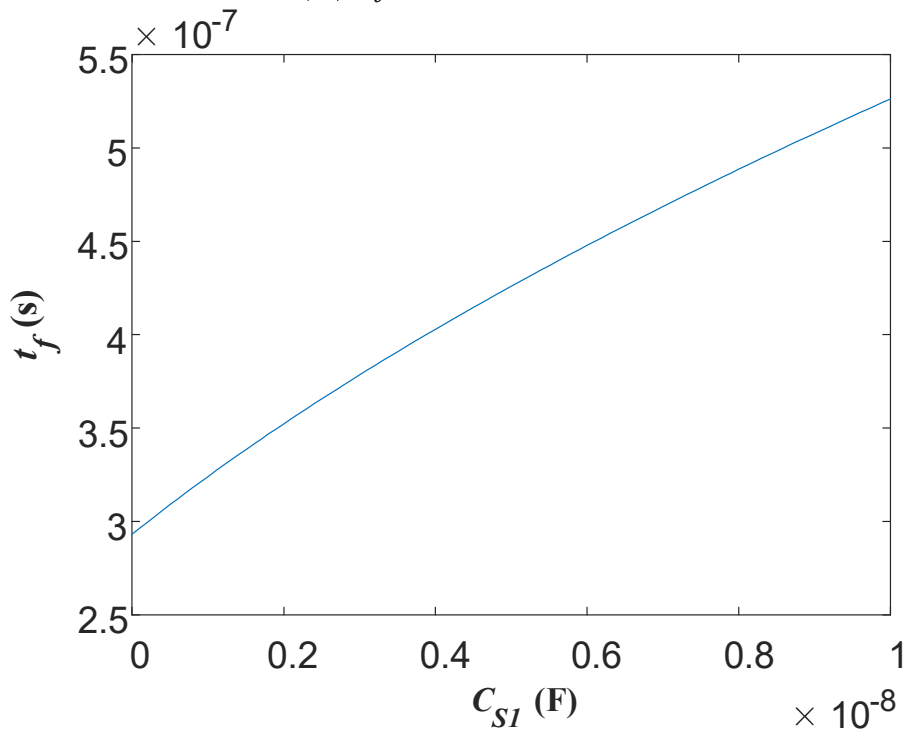

(c) $t_f$ versus $C_{S1}$

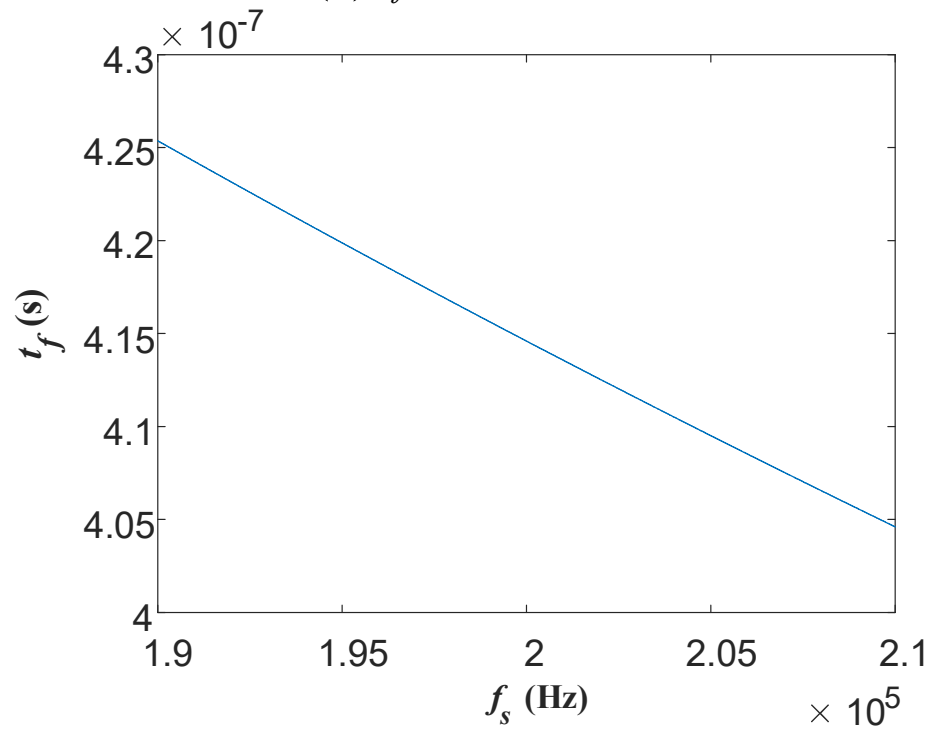

(d) $t_f$ versus $f_s$

Fig. 4. Plots of output voltage $v_o$ versus duty cycle $D$, $t_f$ versus $C_{D1}$, $t_f$ versus $C_{S1}$ and $t_f$ versus $f_s$.

Fig. 4 shows plot of $v_o$ against $D$ based on (12) and plots of $t_f$ versus $C_{D1}$, $C_{S1}$, and $t_f$. Clearly, $v_o$ can be regulated by adjusting the variable $D$. It is observed that the maximum voltage is achieved if $D = 0.5 - f_st_f$ for a given value of $R$. The fact that the output voltage is highly dependent on $R$ implies that a closed-loop system is needed in order to achieve tight regulation of the output voltage. The value of $t_f$ increases as $C_{D1}$ and $C_{S1}$ increased, while $t_f$ drops as $f_s$ increased.

Fig. 5 shows the relationship between the output voltage ($v_o$) and the duty ratio ($D$) with and without resonant capacitors ($C_{S1}$, $C_{D1}$). The latter is referred to as the ideal case since an ideal switch (or diode) contains no parasitic capacitances, i.e., $C_{S1} = 0$ (or $C_{D1} = 0$). By comparing the two curves in Fig. 5, the output voltage ($v_o$) with resonant capacitors (i.e., the bottom curve) is reduced by $1/\pi\,|\,I_{Ls}\,|\,R_{max}\times(1 - \cos(2\pi f_st_f))$, compared with that without resonant capacitors (i.e., the top curve). In addition, the minimum and maximum value of $D$ is shifted by $f_st_f$ and $2f_st_f$, respectively, as illustrated in Fig. 5.

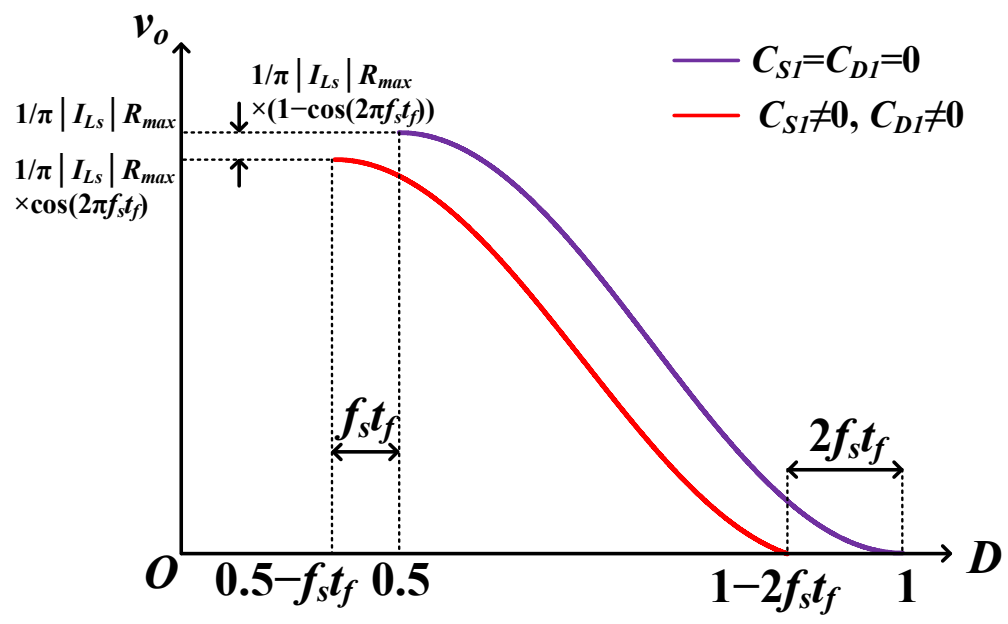

Fig. 5. Plots of output voltage ($v_o$) versus duty ratio ($D$) with and without resonant capacitors ($C_{S1}$, $C_{D1}$) at the maximum load $R_{max}$.

## III. Design Considerations and Implementation

### *A. Design of Reactive Components*

*Design of $L_s$ and $C_s$.* The design objective of $L_s$ is that the quality factor $Q_{Ls}$ of the coil is higher than the desired quality factor $Q$, i.e. $Q_{Ls} \geq Q$. Given that the equivalent series resistance (ESR) of $L_s$ is $R_{Ls\text{-}ESR}$, $Ls$ is designed as

$$Q_{Ls} = \frac{2\pi f_s L_s}{R_{Ls-ESR}} \geq Q \Rightarrow L_s \geq \frac{QR_{Ls-ESR}}{2\pi f_s}. \tag{13}$$

Correspondingly, due to the use of series-series compensation, the capacitor $C_s$ is sized as

$$C_s = \frac{1}{(2\pi f_s)^2 L_s}. \tag{14}$$

*Deign of the Output Capacitor* ($C_o$). The design objective of the output capacitor is to maintain a constant output voltage $v_o$ with sufficiently small voltage ripple $\Delta v_o$, i.e., $\Delta v_o \leq x\% \times v_o$. Fig. 6 provides a graphical illustration of the root cause of the output voltage ripple.

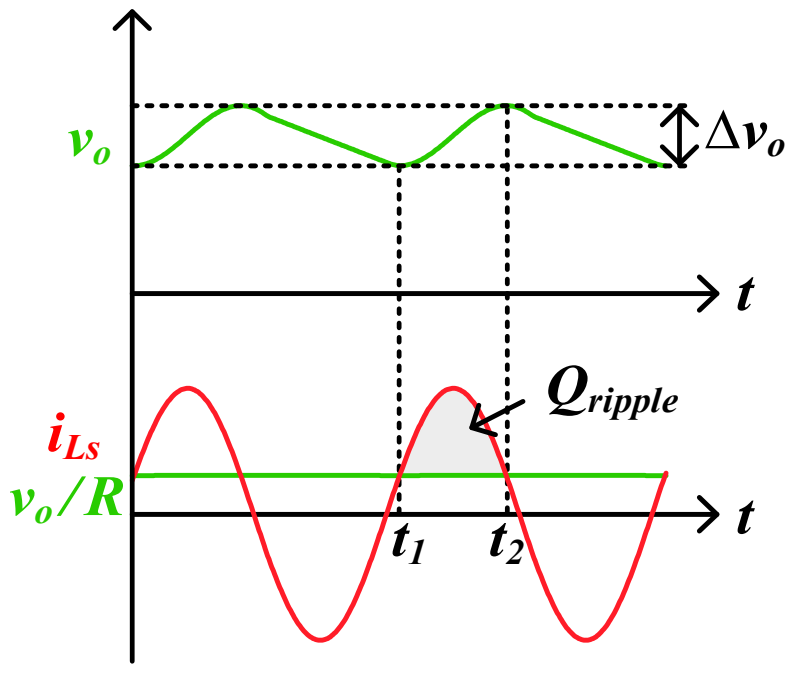

Fig. 6. Illustration of the root cause of the output voltage ripple.

Basically, the rise of the output voltage is attributed to the accumulation of charge $Q_{ripple}$, which is given by

$$\Delta v_o = x\% v_o = \frac{Q_{ripple}}{C_o} \geq \frac{\int_{nT_s}^{(n+0.5)T_s} i_{Ls}\,dt}{C_o} = \frac{|\mathrm{I}_{Ls}|}{\pi f_s C_o}. \tag{15}$$

Hence, the minimum value of the output capacitor is given by

$$C_o \geq \frac{|\mathrm{I}_{Ls}|}{x\% v_o \pi f_s}. \tag{16}$$

*Design of $C_{D1}$ and $C_{S1}$.* Since $C_{D1}$ and $C_{S1}$ are the parasitic capacitance of the diode $D_1$ and switch $S_1$, the values can be obtained from their datasheets.

### *B. Derivation of the Small-Signal Model and Feedback Control for Output Voltage Regulation*

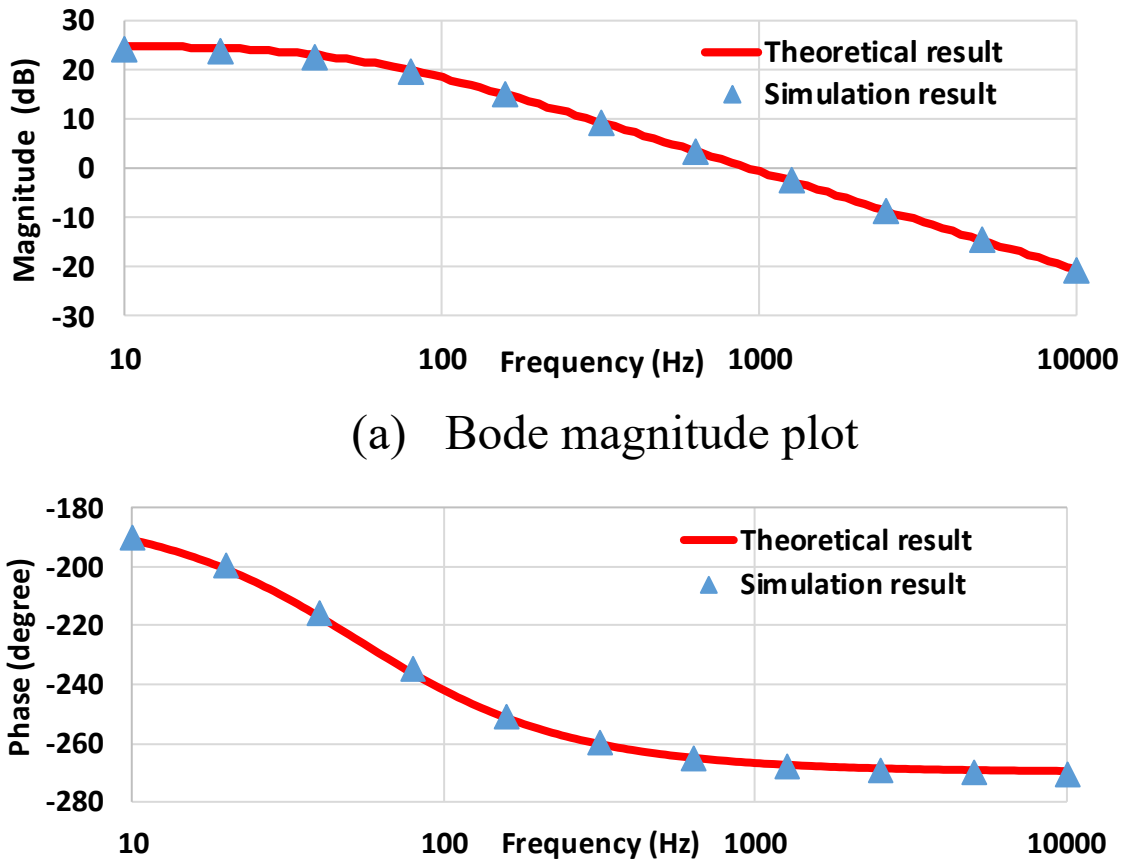

(a) Bode magnitude plot

(b) Bode phase plot

Fig. 7. Theoretical and simulated Bode plots of the open-loop control-to-output voltage using the small-signal model

By linearizing (11) and considering the AC perturbation of duty cycle $D$, the resulting small-signal control-to-output linearized equation can be written as

$$C_O \frac{d\widetilde{v_o}}{dt} = |\mathrm{I}_{Ls}|\widetilde{D}\sin\left(2\pi D + 2\pi f_s t_f\right) - \frac{\widetilde{v_o}}{R}. \tag{17}$$

Fig. 7 shows the Bode plots of the theoretical and simulated small-signal models at $D = 0.5$, $|I_{Ls}| = 1$ A, $R = 30$ Ω, $f_st_f = 0.1$, and $C_o$ = 100 $\mu$F. As is evident in Fig. 7(a) and (b), the theoretical and simulation results are in close agreement, thereby validating the accuracy of the derived small-signal equation from (15). Also, the phase plot in Fig. 7(b) shows that the phase of the open-loop system decreases from −180° to −270°. It is important to note that the use of conventional proportional-integral (PI) controller with positive proportional and integral coefficients ($k_p$ and $k_i$) is unable to provide adequate phase boosting for the uncompensated system, which leads to unstable transient response because of insufficient phase margin and unacceptably large steady-state error due to relatively low DC gain [28]. To address this issue, a modified proportional-integral (PI) controller, as shown in Fig. 8, with negative values of $k_p$ and $k_i$ is employed to attain accurate and stable regulation of the output voltage. In addition, an anti-windup scheme is used to prevent overflow of the integrator and to ensure linear operation of the PI controller. Specifically, the anti-windup loop is added to avoid integrator wind-up, which can occur when the duty ratio ($D$) is saturated. In other words, it prevents the actual value of $D$ from exceeding beyond its upper and lower limits, as defined in equation (6).

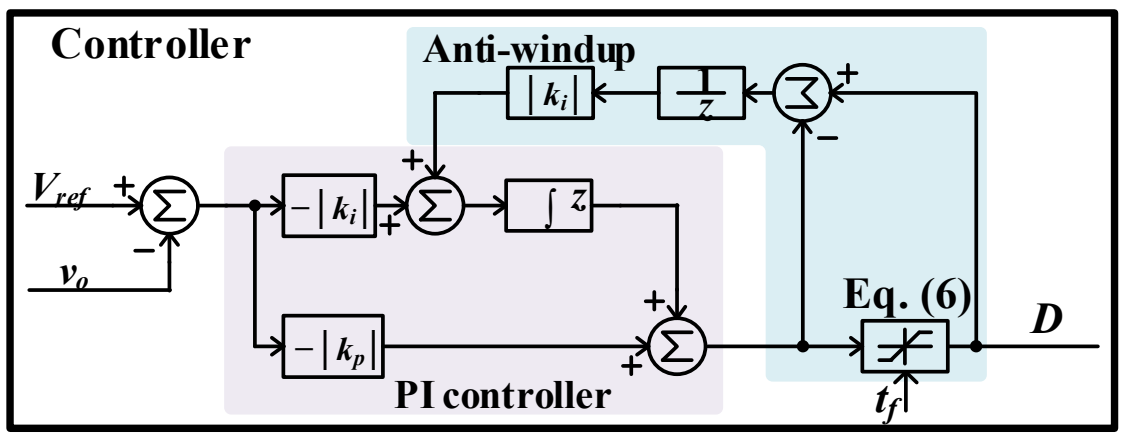

Fig. 8. Block diagram of the modified PI controller with anti-windup.

The proper values of the proportional gain $k_p$ and integral gain $k_i$ of the PI compensator can be determined as follows.

$$k_p = \frac{2\pi f_c C_O}{|I_{Ls}|\sin(2\pi D + 2\pi f_s t_f)}, k_i = \frac{k_p}{RC_O}. \quad (18)$$

where $f_c$ is the desired crossover frequency. Fig. 9 shows the resulting Bode plots of the open-loop transfer function and the closed-loop transfer function of the system with $k_p = -1.07$ and $k_i = -356$. The numerical results of the closed-loop transfer function show that the crossover frequency is at 1 kHz with a phase margin of 90° and a low-frequency gain (at 10 Hz) is 40 dB. The phase margin of 90° ensures a stable closed-loop response and the static gain at DC is high enough (> 40 dB) to eliminate the steady-state error.

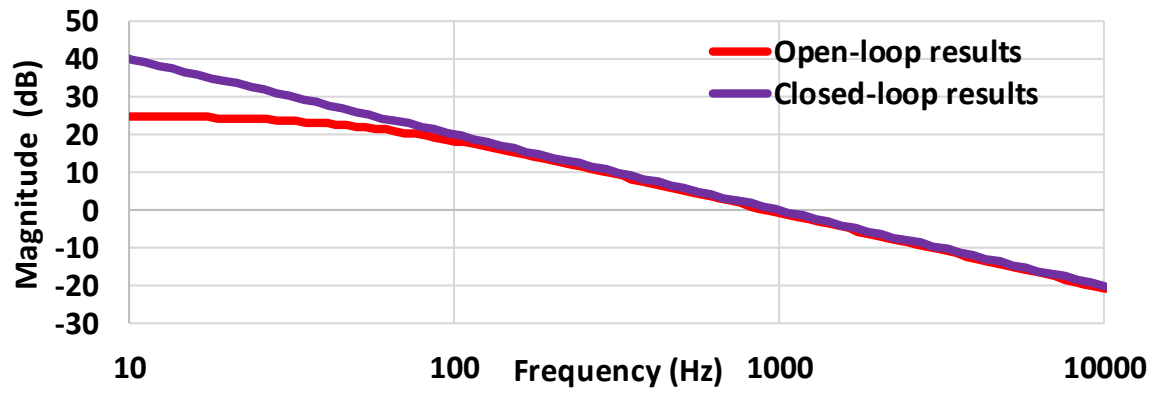

(a) Bode magnitude plot

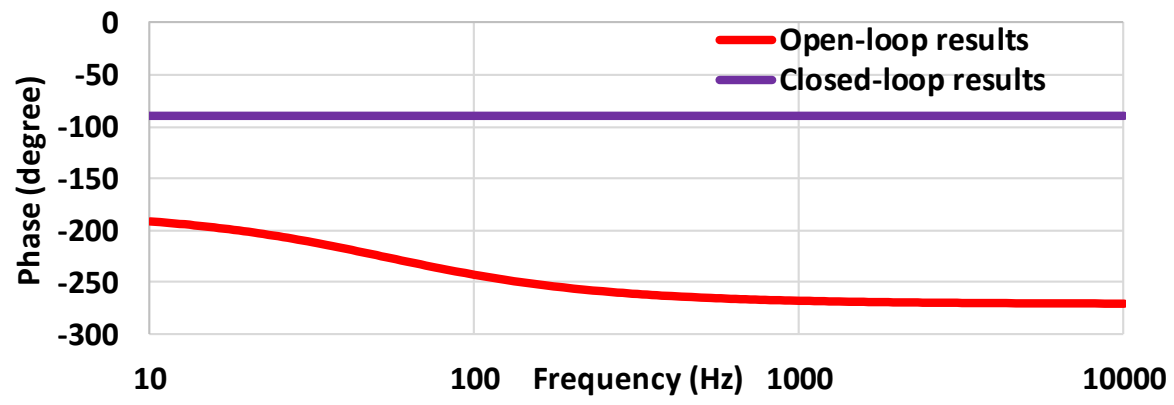

(b) Bode phase plot

Fig. 9. (a) Bode magnitude plot and (b) bode phase plot of the open-loop and closed-loop transfer function.

*C. Synchronization and Hybrid modulation*

Fig. 10 shows the functional block diagram of the frequency synchronization and hybrid modulation scheme. The frequency synchronization is implemented by using the PWM1 module whereas the hybrid modulation is realized by using the PWM2 module of the microcontroller (part number: TMS320F28335). Fig. 11 shows the ideal waveforms of the key signals in frequency synchronization and hybrid modulation.

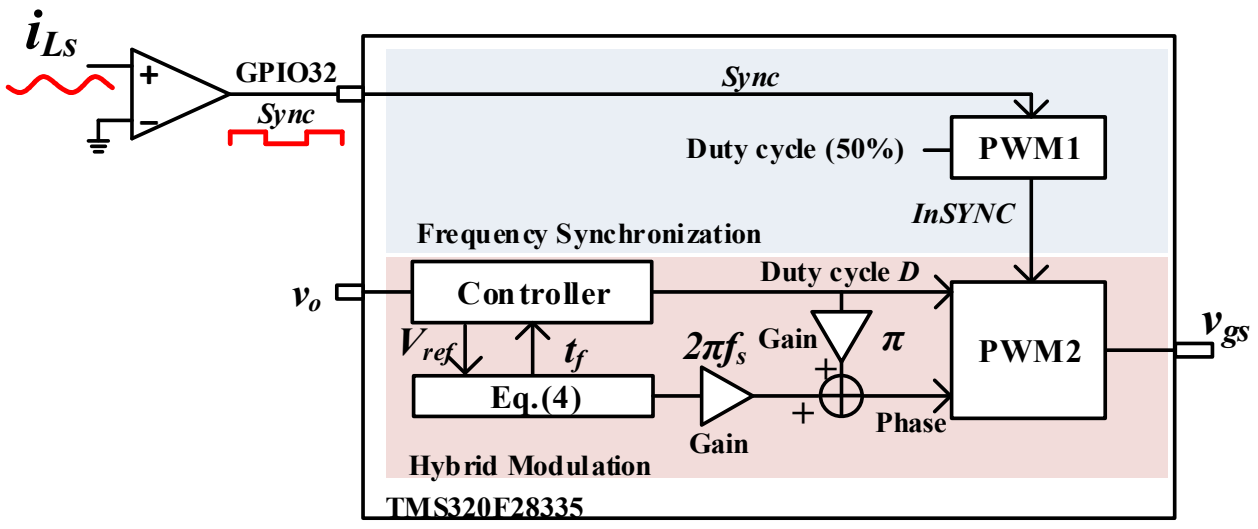

Fig. 10. Block diagram of the frequency synchronization and hybrid modulation.

As can be seen in Fig. 10, an external comparator, which is part of the synchronization circuit, is employed to detect the zero-crossing points of $i_{Ls}$ and generate the corresponding square waveform labelled *Sync* [see Fig. 11].

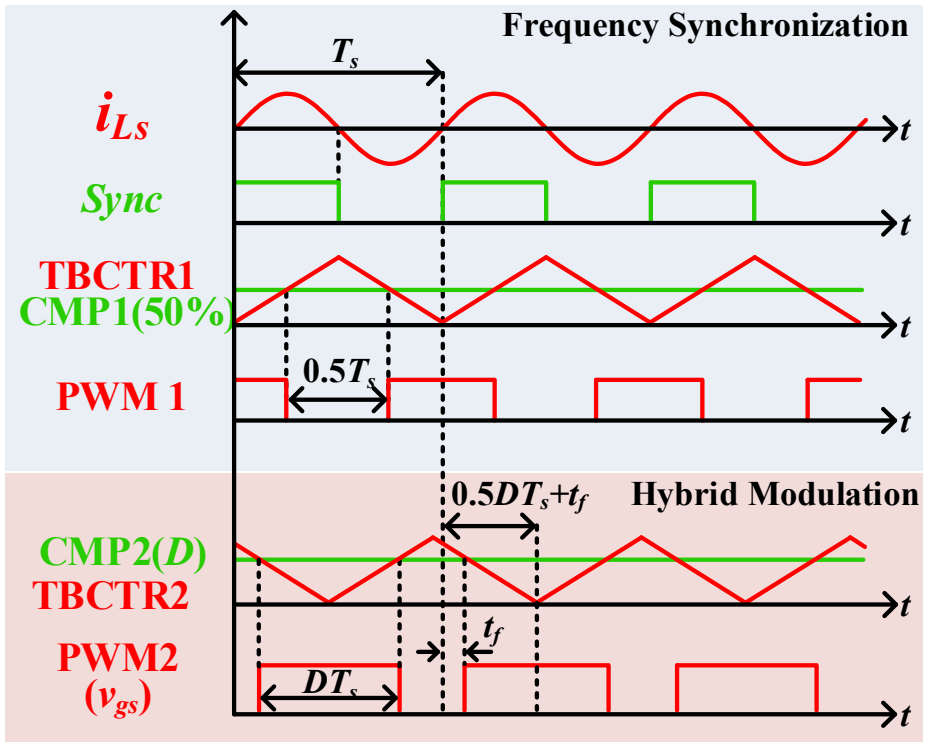

Fig. 11. Ideal waveforms of the key signals in frequency synchronization and hybrid modulation.

The MCU receives this external synchronization signal *Sync* via the GPIO32 pin, which is then used for EPWM synchronization. The rising edge of *Sync* triggers the counter of the PWM1 (TBCTR1) by counting from zero on a cycle-by-cycle basis, which leads to a triangular carrier waveform for PWM1. The duty cycle of the PWM1 is set to be at 50%. Consequently, PWM1 is synchronized with $i_{Ls}$ with a constant phase difference of $0.25T_s$ between them. In this way, frequency synchronization between $i_{LS}$ and $D$ is realized.

Subsequently, hybrid modulation is implemented on the PWM2 module using the PWM1 as a reference. The duty cycle $D$, which is obtained from the controller output [see Fig. 10], is used to compare with the counter of the PWM2 module TBCTR2 in order to adjust the pulse width of the PWM2. On top of them, the phase-shift modulation is realized by adjusting the phase difference between PWM1 and PWM2 (based on the internal synchronization signal *InSYNC*). The desired time delay $t_f$ is calculated from equation (4). To reduce the computation complexity, the real-time output voltage $v_o$ and $|I_{Ls}|$ in equation (4) are replaced by a fixed output voltage reference $V_{ref}$ and a constant value, respectively. Since the carrier waveform TBCTR2 is triangular in shape, the center of the resulting PWM2 is thus aligned with the valley of TBCTR2. Hence, to produce a phase shift of $t_f$, TBCTR2 is simply shifted by $0.5DT_s+t_f$. After performing normalization of the phase angle $\varphi$ of PWM2 by $2\pi$, the effective value of $\varphi$ is given by

$$\varphi = 2\pi f_s t_f + D\pi. \tag{19}$$

Fig. 10 shows the implementation of (17) in the MCU.

## IV. Comparative Study

Table I compares the proposed receiver with the prior art of single-power-switch solutions [16], [17], [22] in terms of topologies, compensation of receiver coil, resonant frequency, etc. As clearly shown in Table I, the proposed receiver uses only one power switch and achieves fully soft-switching operation on the switch and diode. The maximum AC-DC conversion efficiency reaches 98%, which is higher than its predecessors. The regulation frequency is identical to the resonant frequency, which eliminates the use of bulky reactive components. The switch-diode bridge structure can prevent the direct short circuit of the output terminals, thus enhancing the system reliability. Indeed, these features make it a very competitive solution for future WPT applications [3]. Compared with the existing class-E-based solutions [16], [17], a significant advantage of the proposed receiver is that it has relatively low voltage stresses on the power switch. Even though the proposed solution requires an additional diode when compared with [17], the relatively low voltage stresses on both the switch and diode can justify the slight increase in the cost. Compared with the semi-active class-D with pulse density modulation [22], the benefit of the proposed receiver is that it can achieve ZVS turn-on for the power switch, which results in higher efficiency and much higher output power. Additionally, the proposed solution allows the use of a much smaller output capacitance due to the relatively high regulation frequency.

TABLE I. COMPARISON WITH EXISTING WPT RECEIVERS.

| | **Proposed** | [16] | [17] | [22] |
|---|---|---|---|---|
| Topology | **Semi-active Class D** | Modified Class E | Active Class E | Semi-active Class D |
| Compensation of receiver coil | **Series compensation** | Series compensation | Series compensation | Series compensation |
| Resonant frequency | **200 kHz** | 6.78 MHz | 200 kHz | 1 MHz |
| Regulation frequency | **200 kHz** | 6.78 MHz | 200 kHz | 50 kHz |
| Number of power switches | **1** | 1 | 1 | 1 |
| Number of power diode | **1** | 1 | 0 | 1 |
| Maximum Voltage stress | **$V_o$** | ≈$3V_o$ | ≈$3V_o$ | $V_o$ |
| Modulation | **Hybrid modulation** | PWM | Phase-shift Modulation | Pulse Density Modulation |
| Implementation ease | **Yes** | No | No | Yes |
| Soft switching operation | **ZVS for switch ZCS for diode** | ZVS for switch ZCS for diode | ZVS for switch | ZCS for switch and diode |
| Output voltage, power | **24 V, 16 W** | 10 V, 17 W | 24 V, 16 W | 3.1 V, 96 μW |
| Maximum Efficiency | **98 % (Power stage)** | 92 % (Power stage) | 93 % (Power stage) | N/A |

TABLE II PARAMETERS OF COMPONENTS

| Part | Value |
|---|---|
| $C_{D1}$, $C_{S1}$ | 4.5 nF |
| $L_s$ | 172 μH |
| $C_s$ | 3300 pF + 330 pF |
| $C_o$ | 1000 μF |

TABLE III. PART NUMBER OF COMPONENTS.

| Part | Part Number |
|---|---|
| $C_{D1}$, $C_{S1}$ | Parasitics of TK56A12N1 |
| $L_s$ | Custom-made circular copper air coil diameter = 29 cm |
| $C_s$ | B32682A7332K000 (3300 pF) PHE448SB3330JR06 (330 pF) |
| $C_o$ | UVZ1H102MHD |
| **Gate Driver** | ADuM3223 |
| **Comparator** | LM 393P |
| **Current Transformer** | AS-100 |
| **MOSFET $S_1$** | TK56A12N1 |
| **Diode $D_1$** | Body diode of TK56A12N1 |
| **Microcontroller** | TMS320F28335 |

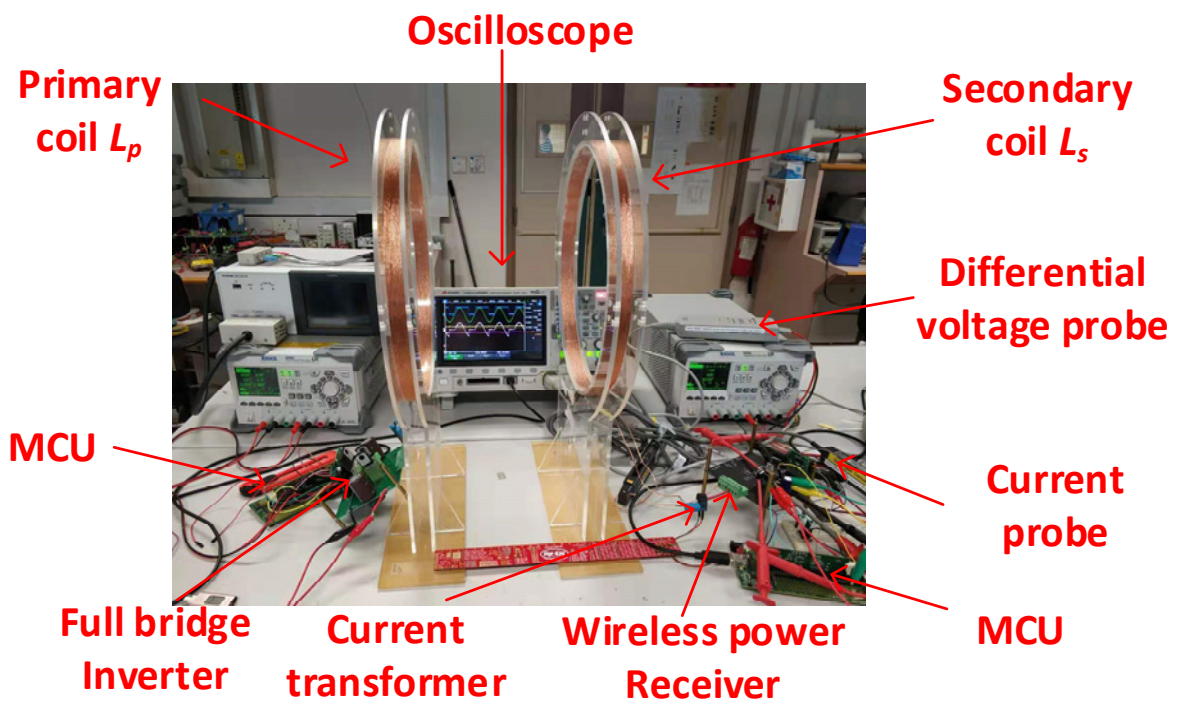

Fig. 12. Experimental setup (including the prototype of the proposed receiver).

## V. Experimental Results

A hardware prototype of the proposed WPT receiver with a switching frequency of 200 kHz switching frequency is constructed for experimental verification. The nominal output voltage is 24 V and the maximum power is 16 W. Fig. 12 shows a photo of the whole experimental setup including the prototype, in which the DP832 power suppliers, DSOX3204T oscilloscope, N2790A differential voltage probe, and 1147B current probe are used. Table II lists the design parameters and part numbers of the key components used in the prototype.

### *A. Steady-State Performance*

Fig. 13 (a) and (b) shows the key waveforms of the proposed receiver in steady-state condition with $v_o$ = 24 V and $i_o$ = 0.63 A. The equivalent load resistance $R$ is 38.09 Ω. The off-state time interval of $v_{CD1}$ and $v_{CS1}$ are measured to be around 2.66 µs and 2 µs, respectively. Hence the duty cycle $D$ is around 0.532. The current of the coil $i_{Ls}$ lead 0.5π against compensated capacitor voltage $v_{Cs}$. The voltage stress of the diode is measured to be 24 V while the measured peak-to-peak input current is 4.7 A. As a sanity check, by substituting the above values of $D$, $t_f$, $i_{Ls}$, and $R$ into (13), the resulting theoretical output voltage is 24.56 V, which agrees very closely with the measured value of 24 V.

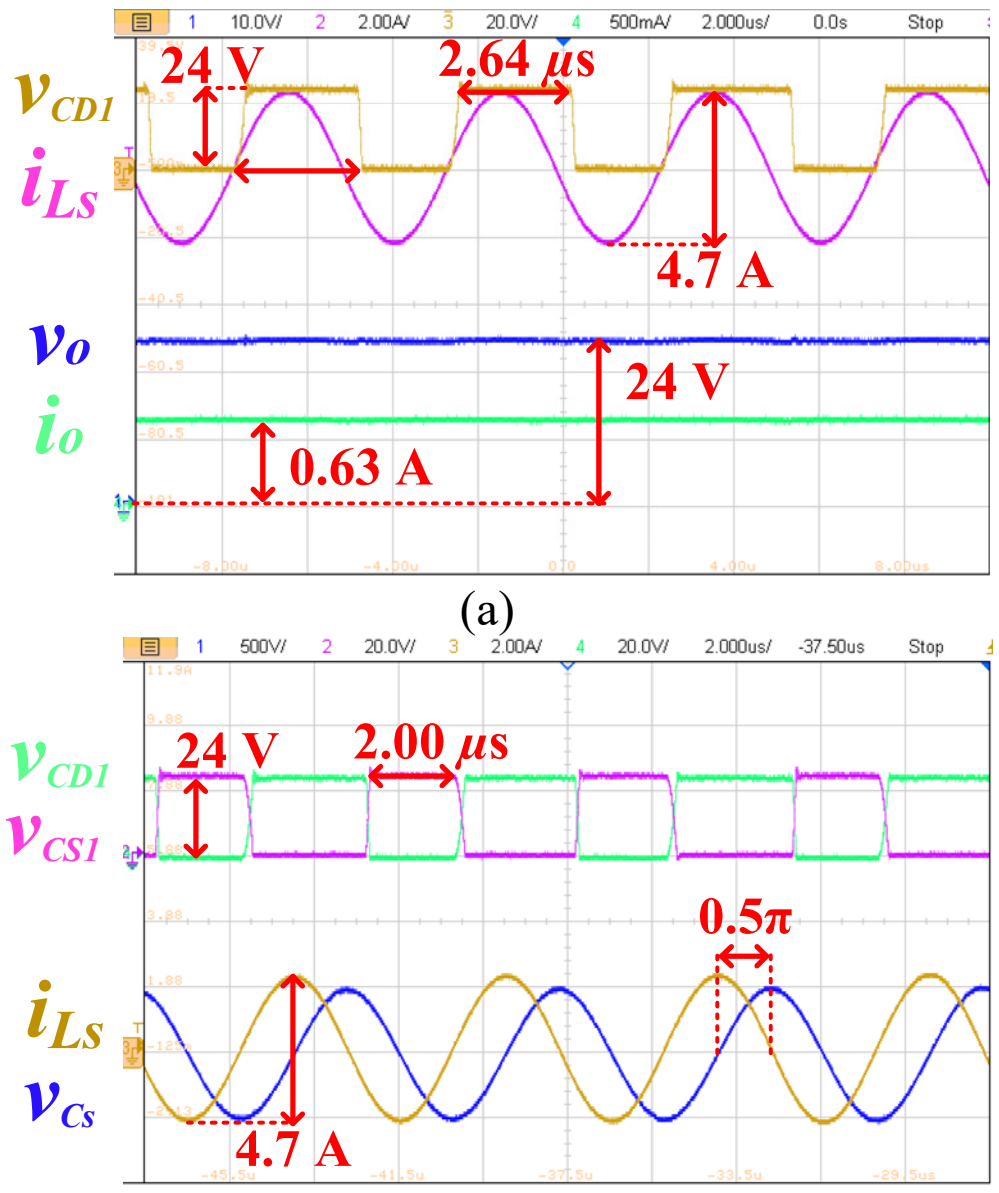

Fig. 13. Steady-state waveforms of $v_{CD1}$, $i_{Ls}$, $v_o$, $v_{Cs}$, $v_{CS1}$ and $i_o$ at a rated output power of 15 W.

Fig. 14 shows the Fast Fourier Transformation (FFT) and THD analysis of the measured waveforms of $v_{CD1}$ and $i_{Ls}$. The fundamental current and THD of the resonant current $i_{Ls}$ are 2.27 A and 2.33%. The high-order (3rd, 5th, 7th, …) harmonic components are suppressed by the wireless power coils. The fundamental voltage and THD of $v_{CD1}$ are 15.02 V and 46.30%, respectively. Since the duty cycle is greater than 50%, the even-order harmonic components are produced, thereby degrading the THD performance. Since $i_{Ls}$ contains very limited harmonic components, the assumption that $i_{Ls}$ is purely sinusoidal in the theoretical analysis remains valid.

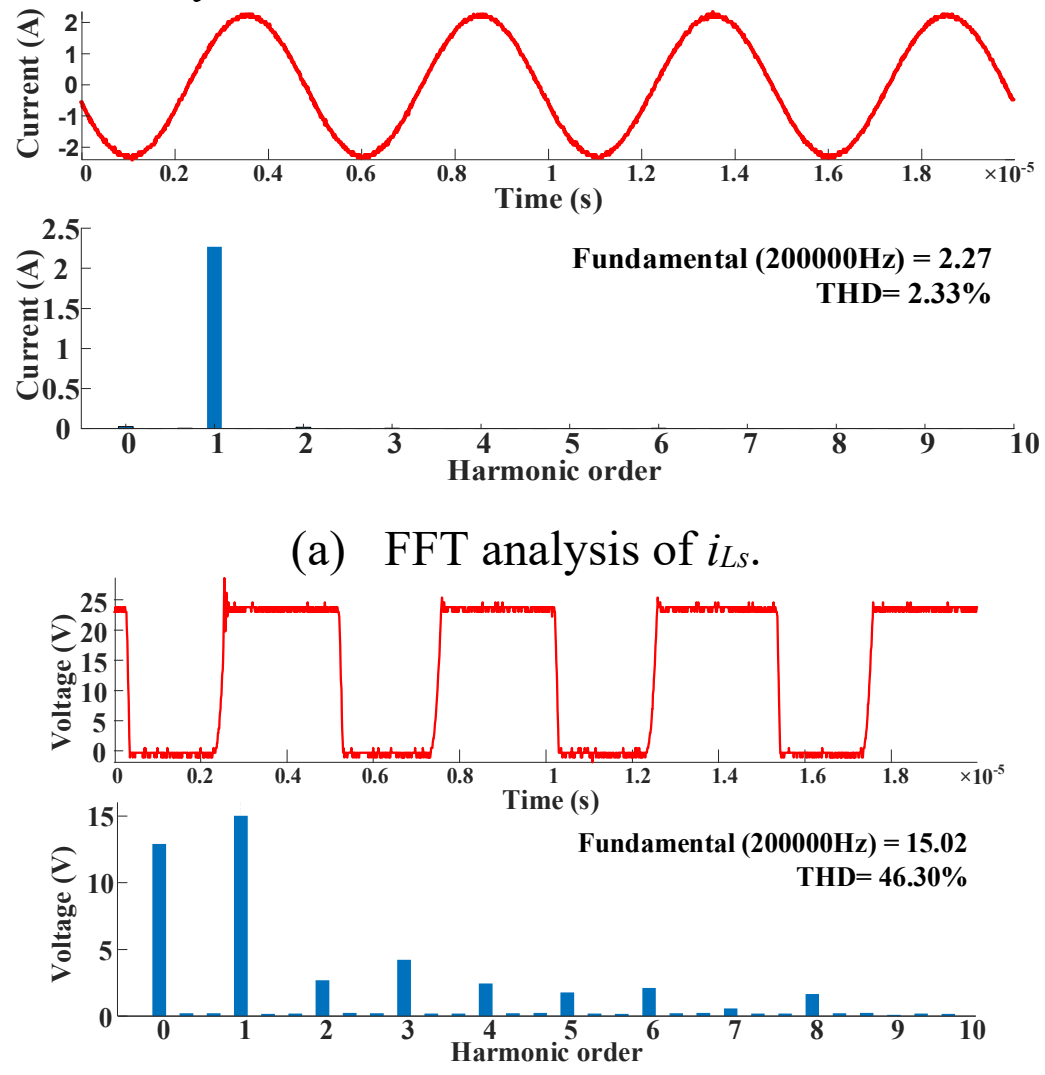

(b) FFT analysis of $v_{CD1}$.

Fig. 14. FFT analysis of the key AC waveforms.

Fig. 15 shows the ZVS operation of the switch and ZCS operation of the diode. Fig. 15(a) shows the turn-on transition of the switch. As soon as $i_{Ls}$ enters in its positive cycle, the voltage of the switch $v_{CS1}$ starts dropping to zero. After $v_{CS1}$ reaches zero, the rising edge of $v_{gs}$ is applied to the switch which turns it on completely. Hence, the switch is turned on with ZVS. The measured falling time $t_f$ is 336 ns. As a sanity check, by substituting the corresponding values of $v_o$, $i_{Ls}$, $C_{D1}$, and $C_{S1}$ into Eq. (4), the theoretical value is 382 ns, which is in good agreement with the measured value. Fig. 15(b) shows the ZCS turn-off transition of the diode. At the end of the conduction state of the diode, $i_{Ls}$, which flows through the diode, reaches zero. Hence, the diode is turned off automatically with ZCS. After a lapse of $t_f$ = 336 ns, $v_{CD1}$ finally reaches 24 V.

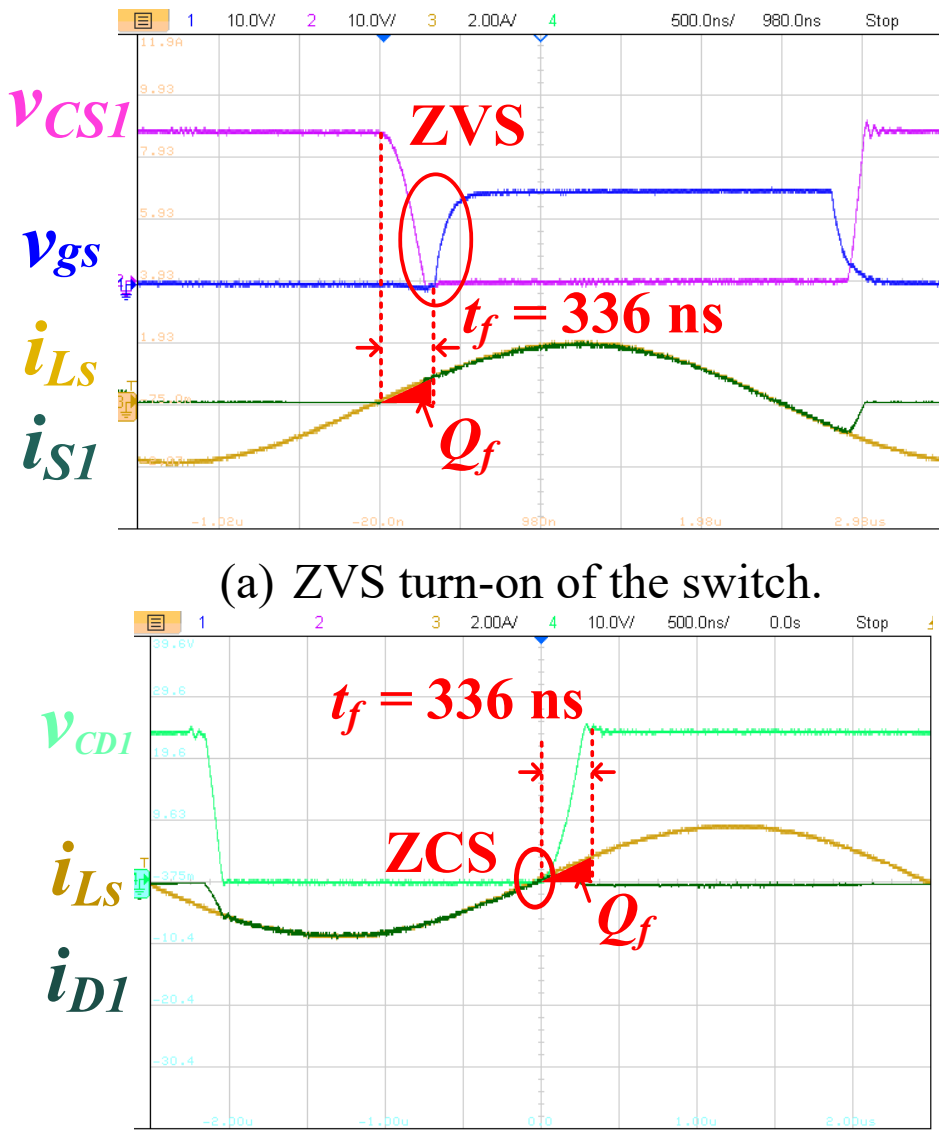

(b) ZCS turn-off of the diode.

Fig. 15. Soft switching operations of the switch and diode.

Fig. 16 shows the measured output voltage and efficiency values of the rectifier power circuit across different output power under output voltage regulation. The nominal output voltage is 24 V. The maximum steady-state error of the regulated output voltage is only 0.1 V, which is around 0.43% of the reference voltage as the power varies from 20% load power to full power. The maximum efficiency of the converter is around 98%, which is achieved at full load condition. At light load condition, the efficiency only reduces slightly to 94%.

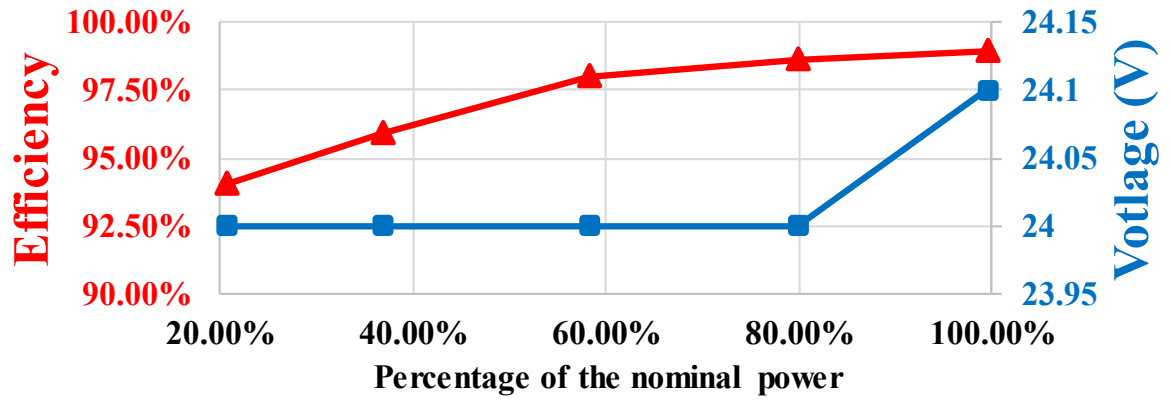

Fig. 16. Measured efficiency and output voltage of the proposed receiver system.

Fig. 17 shows the measured output voltage $v_o$ and amplitude of $i_{Ls}$ of power stage versus coils' distance with output voltage regulation and an output power of 10 W. As the separation distance between the primary and secondary coils increases from 10.5 cm to 21.5 cm, the amplitude of $i_{Ls}$ increases from 1.45 A (at 10.5 cm) to 2.6 A (at 21.5 cm). The maximum voltage regulation error is 0.1 V (which is 0.43% of the reference voltage). Hence, the experimental results in either Fig. 16 or Fig. 17 indicate that the steady-state output voltage regulation error is relatively small over a wide line or load range.

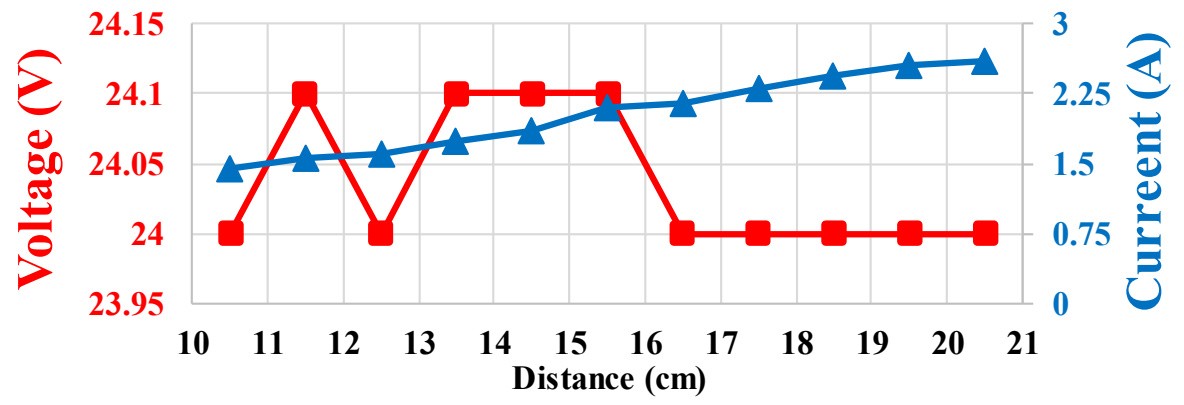

Fig. 17. Output voltage $v_o$ and amplitude of $i_{Ls}$ of power stage versus the coils' separation distance with output voltage regulation.

*B. Dynamic Performance*

Fig. 18 shows the measured waveforms of the rectifier operating with synchronization on/off transition. Prior to the enabling of synchronization, the waveforms of the rectifier have significant fluctuations. As soon as synchronization is enabled, the output voltage ramps up to the nominal value of 24 V within 69 ms. Note that there is no overshoot during the start-up process. Besides, any undesirable low-frequency fluctuations in the waveforms of the rectifier due to frequency asynchrony are eliminated. Therefore, the experimental results demonstrate the effectiveness of the synchronization scheme.

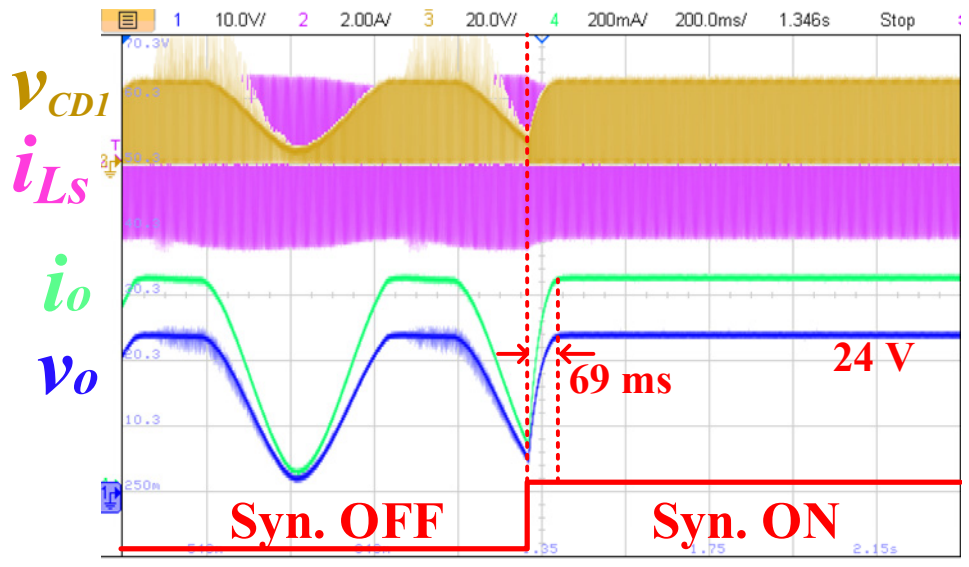

Fig. 18. Measured waveforms of the rectifier operating with synchronization on/off transition.

Fig. 19 (a) and (b) depict the measured waveforms of the rectifier operating with respect to the step-current changes under output voltage regulation. When the output power is increased from 0 W to 16 W, the output voltage experiences a 0.6 V dip (which is 2.5% of the reference output voltage). The settling time for the step-up load change (i.e., from no load to full load) is 8 ms. Conversely, when the output power is reduced from 16 W to 0 W, the resulting overshoot is 0.6 V. The corresponding settling time for the step-down load change is 8 ms.

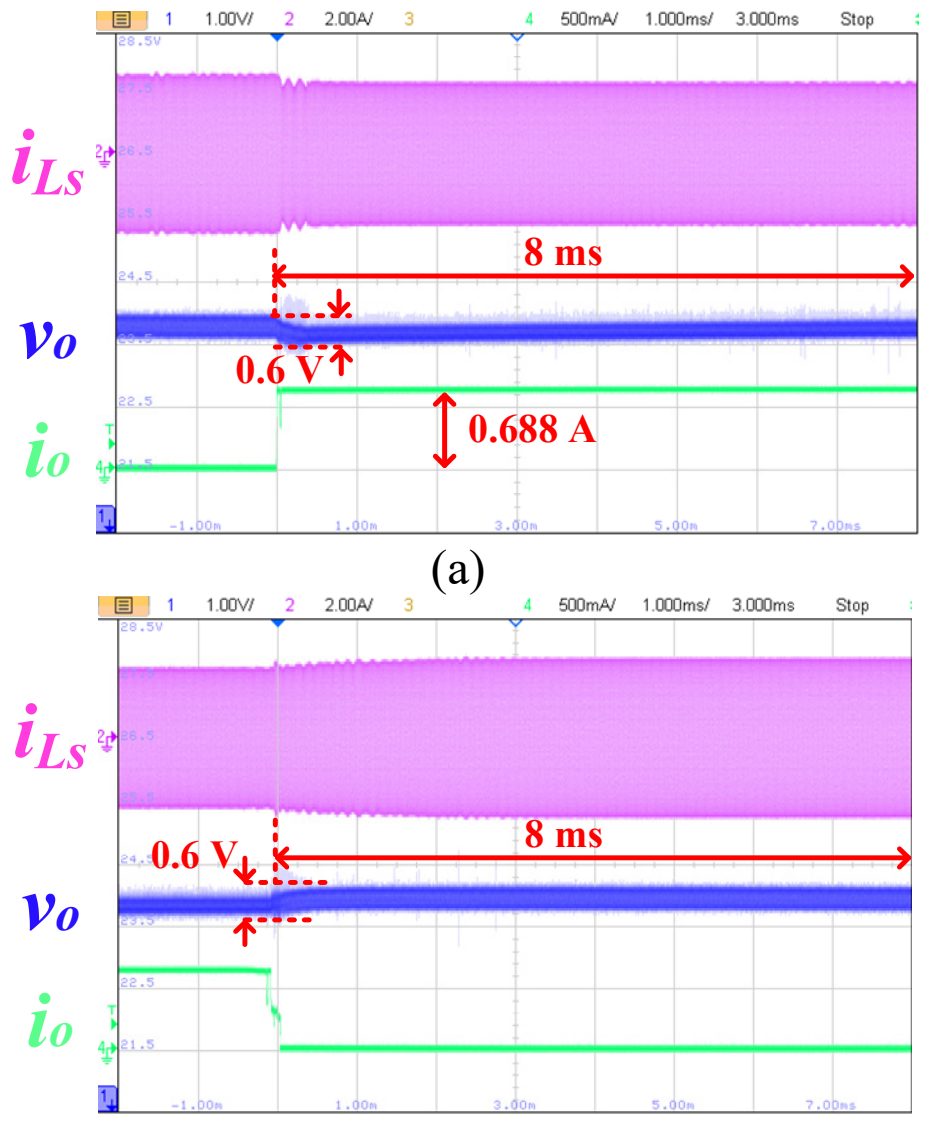

(a)

(b)

Fig. 19. Measured waveforms of the rectifier in response to step changes in the output power with output voltage regulation.

Fig. 20 (a) and (b) show the measured waveforms of the rectifier in response to the input current $i_{Ls}$ changes in light-load condition. When peak-to-peak value of $i_{Ls}$ increases from 2 A to 3.7 A in 10 ms, the output voltage experiences an overshoot of 0.325 V (i.e., 1.35% of the nominal voltage). Likewise, when the peak-to-peak value of $i_{Ls}$ drops from 3.7 A to 2 A in 10 ms, the undershoot is also 0.3 V. The measured dynamic responses corroborate the robustness and stability of the output voltage regulation against variations in load and input current.

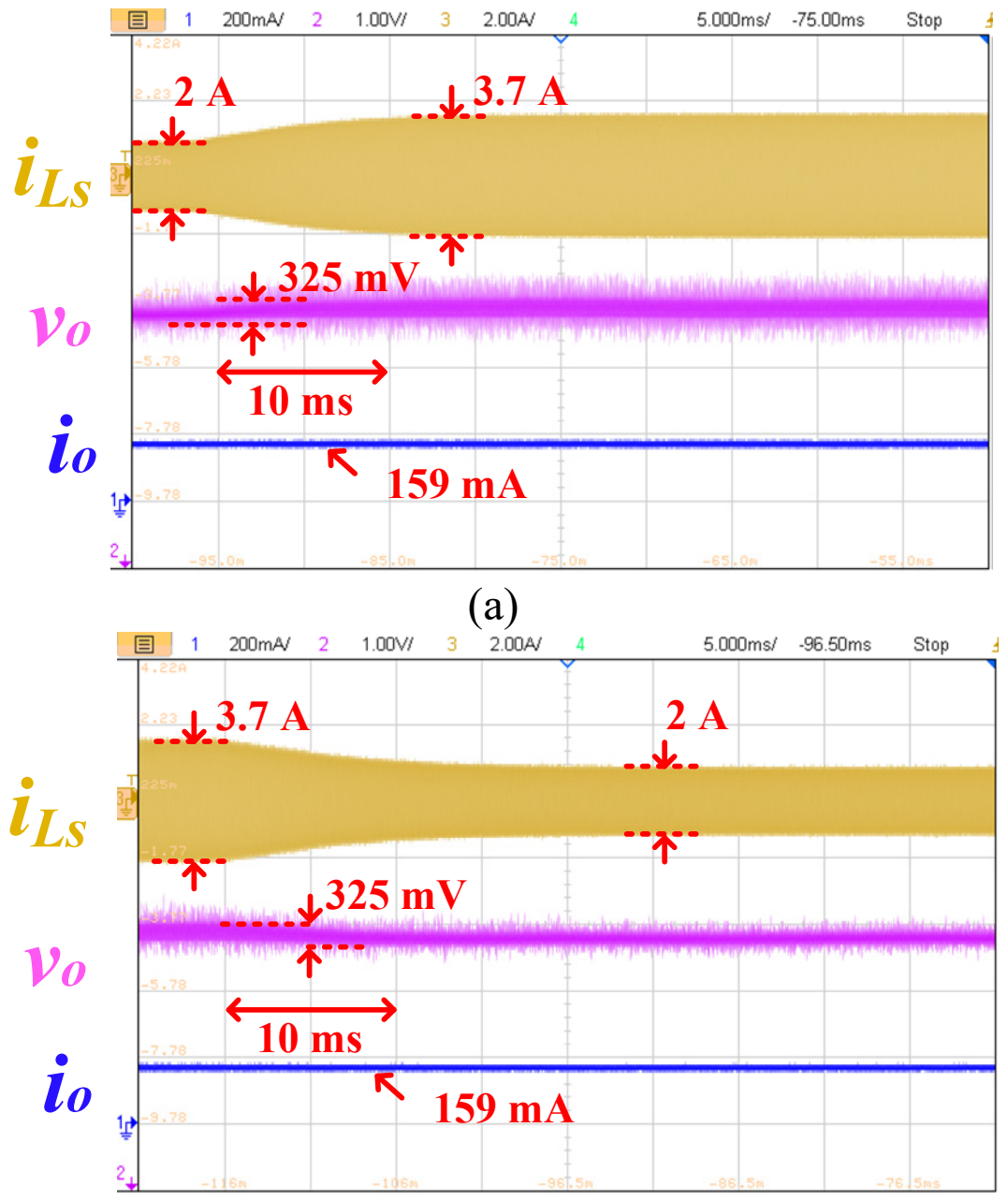

(a)

(b)

Fig. 20. Measured waveforms of the rectifier in response to step changes in the input current with output voltage regulation.

## VI. Conclusions

In this paper, a highly-efficient single-switch-regulated resonant wireless power receiver system with hybrid modulation is presented. The hybrid modulation scheme with phase-shift and pulse width modulations is employed to simultaneously achieve very high efficiency and good output regulation. A comparative study with the existing single-switch wireless power rectifiers highlights the conspicuous advantages of the proposed rectifier, which include low component count, simple design procedure, high efficiency, and cycle-by-cycle output regulation. A hardware prototype of the entire WPT system has been constructed and tested for experimental verification. The experimental results clearly demonstrate that high efficiency and accurate output voltage regulation of the proposed resonant receiver is achievable. The measurement results also confirm good transient responses of the output voltage under output voltage regulation.